\documentclass[11pt]{article}

\usepackage[final]{acl}

\usepackage{times}
\usepackage{latexsym}

\usepackage{graphicx}
\usepackage[table]{xcolor}
\usepackage{array}
\usepackage{makecell}
\usepackage{tcolorbox} 
\usepackage{multirow}
\newcolumntype{C}[1]{>{\centering\arraybackslash}m{#1}}

\usepackage[T1]{fontenc}

\usepackage[utf8]{inputenc}

\usepackage{microtype}

\usepackage{inconsolata}

\newif\ifshowcomments
\showcommentstrue

\ifshowcomments

\newcommand{\authorcomment}[3]{\textcolor{#1}{[\textsf{#2} -- #3]}}
\newcommand{\tomer}[1]{\authorcomment{purple}{#1}{Tomer}}
\newcommand{\amir}[1]{\authorcomment{blue}{#1}{Amir}}
\newcommand{\ori}[1]{\authorcomment{cyan}{#1}{Ori}}

\else

\newcommand{\authorcomment}[3]{}
\newcommand{\tomer}[1]{}
\newcommand{\amir}[1]{}
\newcommand{\ori}[1]{}

\fi

\title{The Overlooked Role of Graded Relevance Thresholds\\in Multilingual Dense Retrieval}

\author{Tomer Wullach \and Ori Shapira \and Amir DN Cohen \\
        OriginAI \\ \texttt{\{tomerw,oris,amirc\}@originai.co}}

\begin{document}
\maketitle
\begin{abstract}
Dense retrieval models are typically fine-tuned with contrastive learning objectives that require binary relevance judgments, even though relevance is inherently graded. We analyze how graded relevance scores and the threshold used to convert them into binary labels affect multilingual dense retrieval. Using a multilingual dataset with LLM-annotated relevance scores, we examine monolingual, multilingual mixture, and cross-lingual retrieval scenarios. Our findings show that the optimal threshold varies systematically across languages and tasks, often reflecting differences in resource level. A well-chosen threshold can improve effectiveness, reduce the amount of fine-tuning data required, and mitigate annotation noise, whereas a poorly chosen one can degrade performance. We argue that graded relevance is a valuable but underutilized signal for dense retrieval, and that threshold calibration should be treated as a principled component of the fine-tuning pipeline.

\end{abstract}

\begin{figure}[t!]
  \centering
  \includegraphics[clip, trim=1.83cm 6.63cm 21.29cm 1.94cm, width=0.8\linewidth]{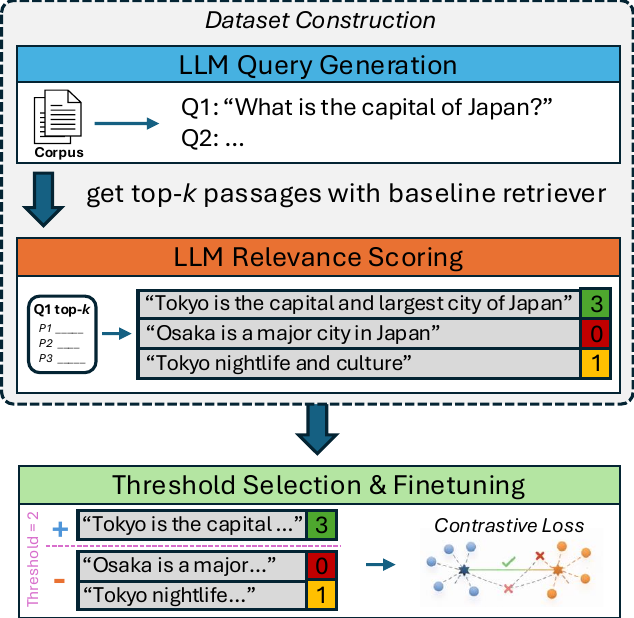}
  \caption{Dense retrieval models are predominantly trained with contrastive learning. Annotating data with graded relevance scores, rather than binary labels, enables selecting an appropriate threshold for positive and negative pairs. This threshold choice has a strong impact on contrastive fine-tuning performance, particularly in multilingual settings.}
  \label{fig:pipeline}
\end{figure}




\section{Introduction}
\label{sec_introduction}
Dense retrieval models have become the standard approach for information retrieval and are now widely used across multilingual search applications \cite{wang2024me5, chen2024bge}. These models are typically fine-tuned with contrastive objectives that rely on distinguishing positive (``relevant'') from negative (``non-relevant'') query-passage pairs \citep{gao-etal-2021-simcse, izacard2022contriever}. Although this supervision is constructed as a binary decision, relevance is rarely binary in practice \citep{Vakkari2004relevancelevels, Sakai2021gradedrelevance}. Passages can vary substantially in how strongly they address a query, and collapsing this variation into a single cutoff removes information that can play a meaningful role during training.

Whether relevance annotations come from human labelers or LLMs, they naturally reflect degrees of relevance. The question is how to leverage that graded structure when the model ultimately trains on binary distinctions. Graded labels are converted to positives and negatives by selecting a threshold on the relevance scale. This step determines which examples the model treats as relevant and which it learns to push apart. Even when graded relevance is available, the choice of where to place the threshold is often overlooked as a significant design decision \citep{Falk2009RelevanceMismatch, Radlinski2010binarizedRelevance}. Yet it shapes the diversity of positive examples, the sharpness of the learned decision boundary, and, as our analyses show, even the amount of fine-tuning data needed to obtain improved performance.

In this paper we examine the role of graded relevance and threshold selection in multilingual dense retrieval. We use an LLM-based pipeline to construct a multilingual dataset with fine-grained relevance labels, as illustrated in \autoref{fig:pipeline}, but our analyses apply equally to datasets annotated by humans or produced through any multi-level relevance scheme. Across three retrieval scenarios: monolingual retrieval (\S\ref{sec_monolingual}), multilingual mixture fine-tuning (\S\ref{sec_mixture}), and cross-lingual retrieval (\S\ref{sec_crosslingual}), we observe patterns that demonstrate the advantages of configuring the threshold.\footnote{Our annotated data and code for analyses will be released.}
The optimal threshold varies across languages and tasks, often aligning with factors such as language resource level, pre-training coverage, and the linguistic match between query and corpus. Lower-resource languages tend to benefit from more inclusive thresholds that preserve broader relevance, while higher-resource languages often prefer stricter thresholds during fine-tuning. When languages are mixed during fine-tuning, threshold interactions grow even more pronounced.

Our analysis also highlights the influence of annotation variability (\S\ref{subsec_monolingual_setup}). We observe disagreement between LLM annotators, especially in lower-resource languages. Thresholding offers a simple way to mitigate the effects of such noise while still making effective use of graded labels.

Taken together, the findings in our analyses support a clear position. Graded relevance provides valuable information for dense retrieval models, but its benefits are realized only when paired with a thoughtfully chosen threshold. Threshold calibration should therefore be viewed as an important component of the fine-tuning pipeline in multilingual settings, one that can improve effectiveness, stabilize training, and reduce the amount of data required for strong retrieval performance.

\section{Background and Related Work}
\label{sec_related_work}

\subsection{Dense Retrieval and Contrastive Learning}

Dense retrieval has become the dominant paradigm in modern information retrieval due to its ability to semantically represent texts \citep{Zhao2024densesurvey}. Models under this paradigm learn query and passage embeddings via contrastive learning, that draws relevant pairs near and separates non-relevant ones \citep{karpukhin2020dense, izacard2022contriever, Hambarde2023irsurvey}.
Although alternative objectives exist, including generative retrieval \citep{Xiaoxi2025genRetrieval} and late-interaction models \citep{Khattab2020colbert}, contrastive learning remains the standard training approach. This motivates our focus on how relevance signals are converted into contrastive supervision.

\subsection{Binary versus Graded Relevance Signals}

Relevance is inherently graded \citep{Saracevic2007relevance}, and has long been treated as such in IR \textit{evaluation}, where metrics like nDCG \citep{Jarvelin2002ndcg} and ERR \citep{Chapelle2009ERR} explicitly model degrees of relevance \citep{Roitero2018scaledrelevance}. In contrast, \textit{training data} for dense retrieval is predominantly binarized. While prior work has explored the benefits of using more fine-grained relevance judgments for training \citep{liu-etal-2021-improving-embedding-based, liu-etal-2021-quadrupletbert, liu-etal-2022-label, tsirigotis2025bixse}, the prevailing practice remains binary labeling, largely for simplicity and scalability.

Several studies point to the limitations of binary relevance labeling. When used for LLM-based annotation, binary prompting can lead to ambiguous or inconsistent judgments because it forces a single decision where gradation is natural \citep{Meng2025llmJudgement}. In contrast, graded labels offer more nuanced supervision \citep{Arabzadeh2025relevancebenchmark}. Human relevance studies echo these findings. \citet{Roitero2018scaledrelevance} show that binary judgments have the lowest agreement with human rankings and emphasize that the appropriate relevance scheme depends on dataset and task requirements.

This body of work highlights that graded relevance is informative and that reducing it to binary form discards useful signals. However, prior work focuses on collecting graded labels rather than understanding how thresholding these labels affects model training, which we address in this paper.

\subsection{Multilingual and Cross-Lingual Retrieval}

Dense retrieval has also advanced multilingual retrieval, supporting shared training across languages and enabling cross-lingual retrieval \citep{suraj2022colbertx, wang2024me5}. Nonetheless, these models exhibit substantial performance disparities across languages, often correlated with resource levels and pre-training exposure. Benchmarks such as MIRACL \citep{Zhang2023miracl} and CLIRMatrix \citep{sun2020clirmatrix} further expose wide variability in multilingual and cross-lingual retrieval quality, shaped by language-specific resource constraints and annotation reliability.

Despite this variability, fine-tuning data in multilingual IR is almost always labeled binarily. This is notable given evidence that multilingual retrieval is especially sensitive to labeling noise. For instance, \citet{shen-etal-2022-recovering} 
show the importance of diverse negative instances,
and \citet{huang-etal-2025-boosting} find that careful negative selection is crucial in low-resource languages. These findings implicitly suggest the value of capturing degrees of relevance, though they do not address threshold calibration directly.

Our analyses extend this work by showing that the optimal relevance threshold depends on language resource level, annotation consistency, and task setting, underscoring the need for flexible threshold calibration in multilingual dense retrieval.

\section{Analysis Setup}
\label{sec_analysis_setup}
In the passage retrieval task, a query is provided as input and the most relevant passages from a corpus are returned~\cite{karpukhin2020dense, chang2020pre}. 
A dense retrieval model produces vector representations (embeddings) for the query and passages, and matches the query with the closest passages, typically measured via cosine similarity.

Dense retrieval models are trained using contrastive learning objectives, which encourage the model to pull positive (``relevant'') query–passage pairs closer in the embedding space while pushing negatives (``irrelevant'') further apart~\cite{gao-etal-2021-simcse, wang2022text}. Accordingly, in order to fine-tune a model for a particular language or domain, an appropriate set of positive and negative examples is required. Increasingly, practitioners rely on large language models (LLMs) to generate or annotate such data~\cite{wang2024improving, lee2024gecko}, enabling scalable and cost-effective dataset construction. We follow such an LLM-based pipeline for producing instances used to fine-tune a dense retrieval model, illustrated in \autoref{fig:pipeline}, which proceeds as follows:

\begin{enumerate}
    \item \textbf{Query generation:} Given a corpus, an LLM is prompted to generate synthetic queries for sampled passages. The generated queries span multiple information-seeking styles, including natural-language questions and keyword-based queries.
    \item \textbf{Candidate retrieval:} For each generated query, a baseline dense retrieval model retrieves the top-$k$ candidate passages from the corpus (some of which may be irrelevant when there are few results).
    \item \textbf{Relevance labeling:} Each query-passage candidate pair is presented to an LLM, which assigns a relevance score on a predefined scale (possibly binary). This ranking often differs from, and is generally more reliable than, the ranking from the previous step, as also reported by \citet{lee2024gecko}.
    \item \textbf{Dataset construction:} The resulting fine-tuning dataset contains instances $(q, p_i, s_i)$, where $q$ is a query, $p_i$ is one of the $k$ retrieved passages, and $s_i$ is the assigned relevance score.
\end{enumerate}

Traditionally, relevance annotation is treated as a binary decision, relevant vs. irrelevant. However, introducing a graded relevance scale can reduce annotation arbitrariness and better reflect the continuum of relevance commonly observed in retrieval tasks (\S\ref{sec_related_work}). Highly similar query-passage pairs are naturally labeled as clearly relevant, e.g.,
\begin{quote}
\small
Q: Health effects of excessive and insufficient apoptosis

P: \textit{``... Excessive apoptosis causes atrophy, whereas an insufficient amount results in uncontrolled cell proliferation...''}
\end{quote}
whereas less obvious but still meaningful matches may warrant intermediate scores rather than being forced into a binary ``irrelevant'', e.g.,
\begin{quote}
\small
Q: Knights and pawns

P: \textit{``Dunsany's Chess... One side has standard chess pieces, and the other side has 32 pawns.''}
\end{quote}

Because contrastive objectives ultimately require binary supervision, a threshold must be applied to convert graded relevance scores into positive and negative instances: scores at or above the threshold become positives, and those below become negatives. Setting this threshold arbitrarily preserves the limitations of purely binary annotation. As the analyses in the subsequent sections show, the threshold choice has consequences.

A higher threshold yields a narrow notion of relevance, exposing the model only to highly similar pairs during training. Lowering the threshold expands the diversity of positive examples, enabling the model to learn more flexible or distant relevance relationships. As we will see, this decision significantly affects performance across languages and tasks.

\begin{figure}[t!]
  \centering
  \includegraphics[clip, trim=0cm 0cm 0cm 0.95cm, width=\linewidth]{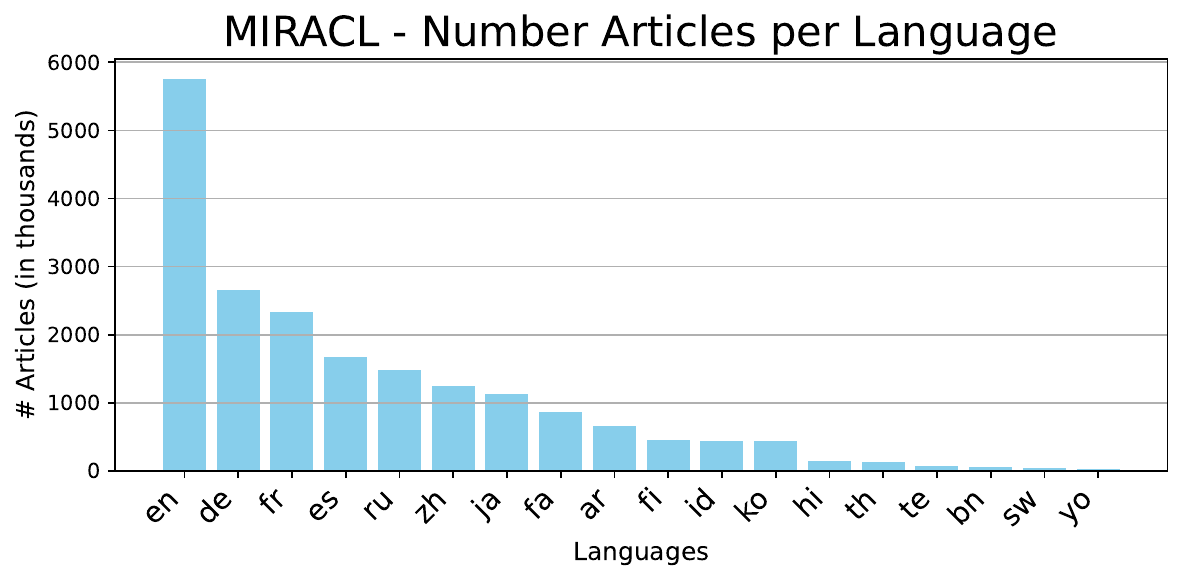}
  \caption{Number of documents (Wikipedia articles) per language in the MIRACL dataset. In our analyses we use Finnish (fi), Arabic (ar), Japanese (ja), Russian (ru), Spanish (es) and English.}
  \label{fig:miracl}
\end{figure}


\subsection{Our Synthetic Dataset}
\label{subsec_synthetic_data_ours}

To support our analyses, we prepared a synthetic fine-tuning dataset with graded relevance scores using the pipeline described above.

\paragraph{Source corpus.}
MIRACL\footnote{\url{https://huggingface.co/datasets/miracl/miracl-corpus}} \citep{Zhang2023miracl} is a large corpus covering 18 languages, used for retrieval tasks. To enable analyses across a range of resource levels, we selected Finnish and Arabic as low-resource languages, Japanese as a medium-resource language, Russian and Spanish as high-resource languages, and English as a very high-resource language. \autoref{fig:miracl} shows MIRACL’s distribution of data by language. For each language, we randomly sampled 500K passages outside commonly used development sets. These six corpora form the basis for our synthetic data.\footnote{These languages were also chosen to align with the test languages available in the cross-lingual evaluation in \S\ref{sec_crosslingual}.}

\paragraph{Query generation.}
We employed \texttt{GPT-4o}\footnote{We used \texttt{GPT-4o} as it offered a practical balance of multilingual capability and cost for large-scale annotation.} \citep{openai2024gpt4ocard} to generate language-matched queries designed to resemble web-search intent. A fixed prompt was used across languages, with only the target-language specification modified (see Appendix \ref{app:implementation}).

\paragraph{Baseline dense retrieval model.}
For candidate retrieval, we used the \texttt{multilingual-e5-large}\footnote{\url{https://huggingface.co/intfloat/multilingual-e5-large}} embedding model \citep[\texttt{me5};][]{wang2024me5} to retrieve a corpus-wide ranking for each query, selecting the top-5 passages. The model is pre-trained using MIRACL and related datasets, and we assume its language familiarity roughly follows the distribution shown in \autoref{fig:miracl}.

\paragraph{Relevance score annotation.}
Each query–passage pair was annotated using \texttt{GPT-4o}, assigning a relevance score from 0 to 3, where 3 denotes high relevance and 0 denotes irrelevance. A four point scale has been shown to be effective in this setting \citep{McDonnell2016relevancejudgments, Roitero2018scaledrelevance}. The prompt and scale definition appear in Appendix \ref{app:implementation}.

\begin{figure*}[t!]
  \centering
  \includegraphics[width=\linewidth]{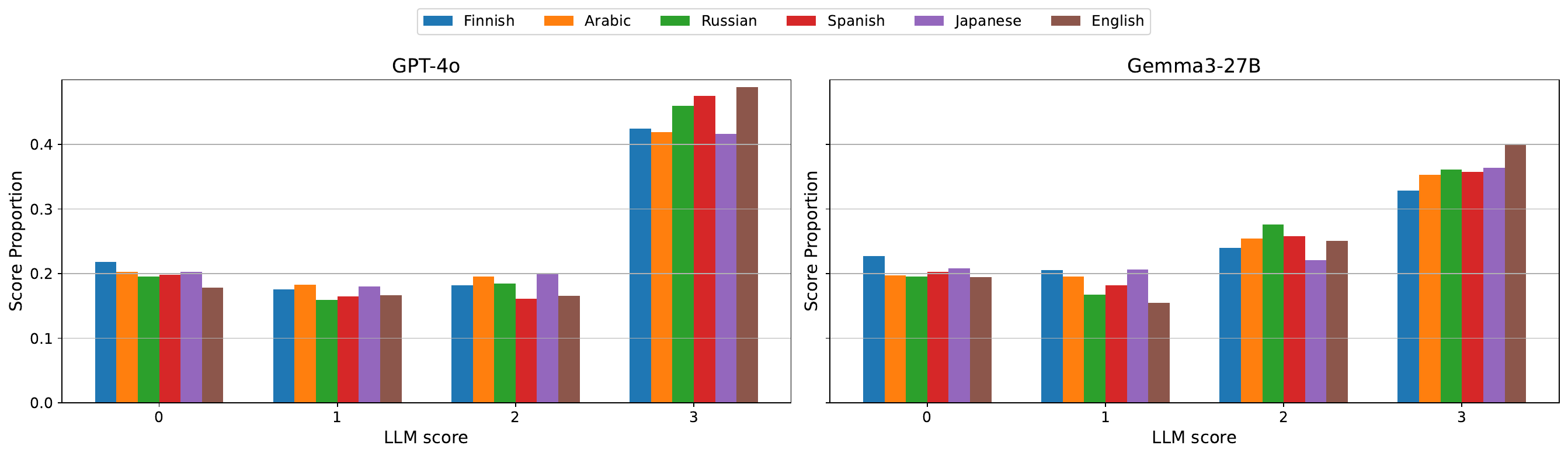}
  \caption{Distributions of relevance scores across six languages, as provided by the two LLM annotators (\texttt{Gpt-4o} and \texttt{Gemma-3-27B}). The relevance score ranges from 0 (passage not relevant to query) to 3 (passage very relevant to query). The main difference is apparant in labels 2 and 3.}
  \label{fig:score_dist}
\end{figure*}

Finally, we sampled down the instances from 500K to 300K, maintaining a similar distribution of scores across the six languages, for fairness in our analyses. \autoref{fig:score_dist} (left side) shows the resulting score distributions, and Appendix \ref{app:implementation} describes the downsampling procedure.

With this dataset, we next examine how different threshold choices affect fine-tuning outcomes. Across three use cases, we argue that graded relevance combined with a thoughtfully chosen threshold provides a more flexible and informative supervision signal than binary annotation alone. Adjusting the threshold allows us to explore the trade-off between specificity and coverage in what the model learns as ``relevant''. Although our analysis focuses on synthetic data, the considerations around graded relevance and threshold calibration apply equally to any fine-tuning dataset, whether curated by humans or by LLMs.

\section{Use-case 1: Monolingual Retrieval}
\label{sec_monolingual}

Monolingual passage retrieval involves retrieving passages in the same language as the query and remains the most common retrieval scenario. Using a multilingual dense retrieval model allows holding architecture and training conditions constant across languages while benefiting from shared multilingual representations. This makes monolingual retrieval a natural first setting for analyzing how fine-tuning with graded relevance affects performance across languages, especially given differences in data availability, annotation quality, and pre-training coverage.

\subsection{Experimental Setup}
\label{subsec_monolingual_setup}

To evaluate the effect of relevance-aware contrastive fine-tuning, we used our dataset annotated with graded relevance scores for query-passage pairs. From this dataset, we constructed several fine-tuning splits by (i) varying the threshold, $\tau\in\{1,2,3\}$, used to designate positive ($s\geq\tau$) and negative ($s<\tau$) pairs, and (ii) varying the number of training instances, from 50K to 300K by intervals of 50K. This allowed us to analyze how both the quantity of supervision and the strictness of the positive/negative boundary influence downstream performance.

We fine-tuned the \texttt{me5} multilingual dense retriever on each configuration, separately for each of the six languages (technical details in Appendix \ref{app:implementation}). The resulting models were evaluated on the respective language subsets of the MIRACL development set, using nDCG@10 as the primary metric.

To additionally assess the effect of the threshold with respect to the relevance annotator, we repeated the annotation and fine-tuning procedure using relevance scores produced by \texttt{Gemma-3-27B} \citep{gemmateam2025gemma3}, a smaller LLM than its counterpart.

\begin{figure*}[t!]
  \centering
  \includegraphics[width=\textwidth]{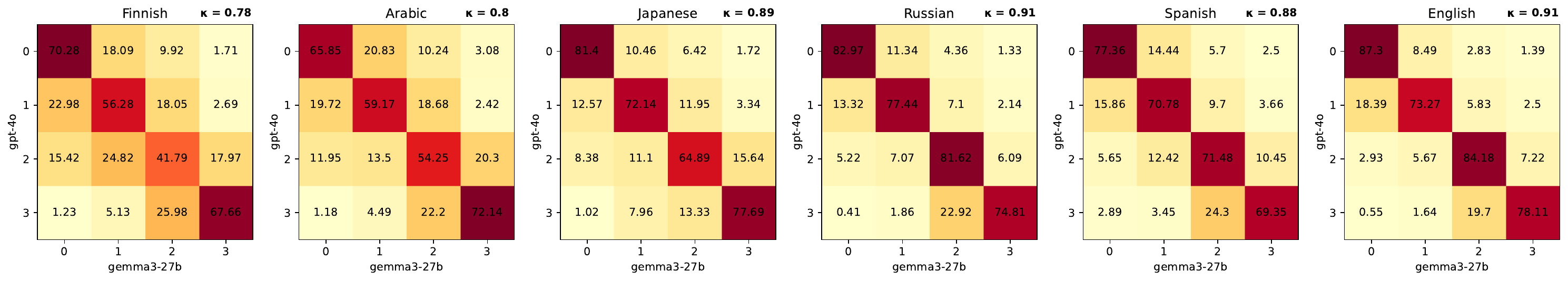}
  \caption{Agreement heatmaps between the two LLM annotators (\texttt{GPT-4o} and \texttt{Gemma-3-27B}) across languages. Each cell shows the row-normalized frequency of \texttt{Gemma-3-27B}’s scores conditioned on \texttt{GPT-4o}’s score. High-resource languages show stronger diagonal agreement (higher Quadratic Weighted~$\kappa$), while low-resource languages exhibit greater off-diagonal noise (lower $\kappa$). These differences in annotation consistency further motivate examining how threshold choices affect the preparation of finetuning data.}
  \label{fig:heatmaps}
\end{figure*}

\paragraph{\texttt{GPT} vs.\ \texttt{Gemma} relevance scores.}
\autoref{fig:score_dist} shows the distributional differences between the two LLM annotators. While the score distributions for classes 0 and 1 are broadly similar, the distributions diverge more strongly for the higher relevance labels. \texttt{GPT-4o} assigns label 3 more readily, whereas \texttt{Gemma-3-27B} tends to prefer label 2 over 3. This may indicate that \texttt{GPT-4o} applies a more permissive criterion for identifying strongly relevant passages, or alternatively that \texttt{Gemma-3-27B} is more conservative when making high-confidence relevance judgments.

\autoref{fig:heatmaps} compares annotator agreement across languages. Agreement is substantially lower for lower-resource languages, mirroring patterns commonly observed in human annotation, where reduced linguistic proficiency leads to noisier judgments. Because LLM abilities vary across languages, particularly for long-tail languages, it is reasonable that relevance scoring becomes less consistent. These differences in annotation consistency further motivate examining how threshold choices affect the preparation of finetuning data. For this use case we include both annotation sources in our experiments and compare the resulting trends.

\begin{figure*}[t!]
  \centering
  \includegraphics[width=\textwidth, , keepaspectratio]{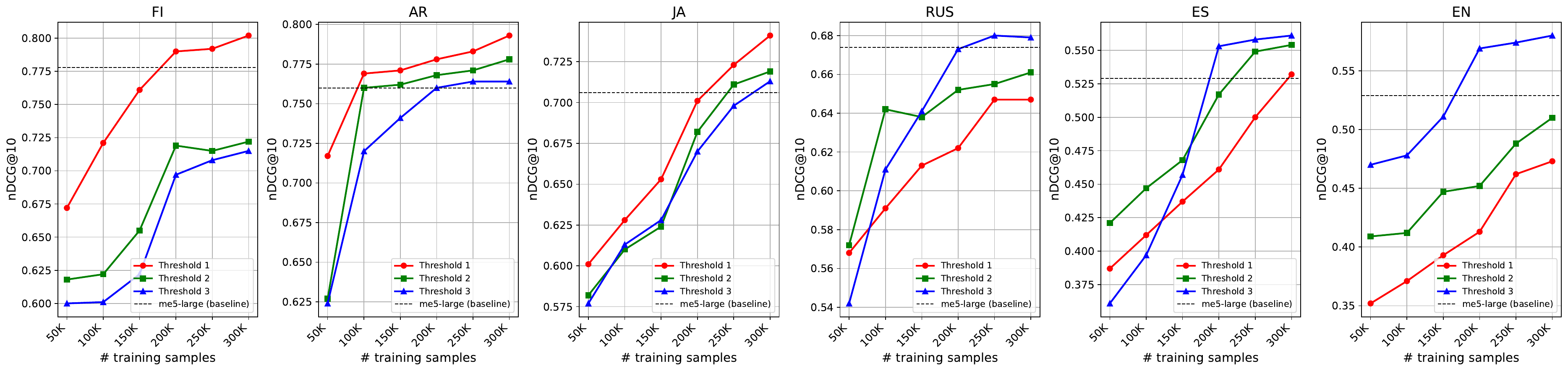}
  \caption{Monolingual retrieval performance (nDCG@10) of \texttt{me5} on MIRACL (horizontal dashed lines) and after fine-tuning, across six languages. Each subplot shows performance as a function of training set size under different relevance thresholds used to convert graded LLM scores into contrastive training pairs. Lower-resource languages tend to benefit from lower thresholds, whereas higher-resource languages perform better with higher thresholds.}
  \label{fig:mono_gpt}
\end{figure*}

\subsection{Results}

\autoref{fig:mono_gpt} presents the results across languages for varying thresholds and data sizes when using \texttt{GPT-4o} as the annotator (\autoref{fig:mono_gemma} in the appendix shows the corresponding \texttt{Gemma-3-27B} results). We find that increasing the amount of fine-tuning data generally improves monolingual retrieval performance. Initially, small training sets cause a noticeable performance drop relative to the unfine-tuned \texttt{me5} baseline, suggesting that insufficient supervision can distort rather than enhance the model’s internal representations. As more annotated pairs are used, performance recovers and eventually begins to plateau.

Thresholding strategy plays a critical role. Models trained with poorly chosen relevance thresholds often fail to match the performance of the baseline model. Perhaps more finetuning data would help surpass the baseline. In contrast, well-calibrated thresholds yield clear performance gains with less finetuning data. This underscores that contrastive fine-tuning requires thoughtful binarization rather than naïve thresholding.

A notable pattern emerges across languages. In lower-resource languages, models benefit from lower thresholds, which preserve a broader diversity of relevance relationships. In higher-resource languages, the opposite trend appears, in which stricter thresholds yield better performance. We hypothesize that this arises from differences in the pre-training corpus. For lower-resource languages, the model likely saw fewer examples and a narrower range of positive signals, making additional diversity during fine-tuning beneficial. For higher-resource languages, however, the model likely already encodes rich and diverse relevance distinctions; injecting overly heterogeneous positives may introduce noise and reduce discriminability.

Differences across languages may also reflect properties of the evaluation data. We observe higher nDCG@10 scores for some lower-resource languages, which may indicate that retrieving relevant passages is comparatively easier, perhaps due to more pattern-based query styles or greater lexical overlap between queries and corpus text. Regardless of whether these effects stem from training data, testing data, linguistic differences, or a combination thereof, the findings collectively highlight the importance of tuning the positive/negative boundary to the target language and domain.

When comparing LLM annotators, we find that models fine-tuned on \texttt{Gemma}-based relevance scores underperform those trained on \texttt{GPT}-based scores. Within the scope of this analysis, we can only speculate about the underlying causes, such as 
semantic sensitivity or language coverage. Importantly, however, the qualitative trends largely hold across annotators, with only a few exceptions, e.g., Spanish slightly preferring a threshold of~2. While thresholding mitigates some annotation noise, the broader point is that variability in relevance judgments adds yet another consideration when selecting an appropriate threshold.

Given the consistently higher performance obtained with GPT-based relevance scores, we adopt these scores for the subsequent two use cases.

\section{Use-case 2: Finetuning with Mixed Languages}
\label{sec_mixture}

\begin{table}[t]
\centering
\small
\setlength{\tabcolsep}{5pt}
\renewcommand{\arraystretch}{1.2}

\resizebox{\columnwidth}{!}{%
\begin{tabular}{|c|c|c|c|c|}
\Xhline{1.2pt}
\textbf{Target lang.} &
\textbf{Additional lang.} &
\textbf{$\tau = 1$} &
\textbf{$\tau = 2$} &
\textbf{$\tau = 3$} \\
\Xhline{1.2pt}

\multirow[c]{5}{*}{\textbf{FI}} &
\textbf{-} & 0.761 & 0.655 & 0.622 \\
\cline{2-5}
& \textbf{AR} & {\scriptsize -0.123 (-16.2\%)} & {\scriptsize -0.064 (-9.8\%)} & {\scriptsize -0.100 (-16.1\%)} \\
\cline{2-5}
& \textbf{RUS} & {\scriptsize +0.049 (+6.43\%)} & {\scriptsize -0.017 (-2.59\%)} & {\scriptsize -0.170 (-27.3\%)} \\
\cline{2-5}
& \textbf{ES} & {\scriptsize +0.056 (+7.4\%)} & {\scriptsize -0.022 (-3.4\%)} & {\scriptsize -0.052 (-8.4\%)} \\
\cline{2-5}
& \textbf{EN} & {\scriptsize +0.070 (+9.2\%)} & {\scriptsize +0.000 (+0.0\%)} & {\scriptsize -0.025 (-4.0\%)} \\
\Xhline{1pt}

\multirow[c]{5}{*}{\textbf{AR}} &
\textbf{-} & 0.771 & 0.762 & 0.741 \\
\cline{2-5}
& \textbf{FI}  & {\scriptsize -0.119 (-15.4\%)} & {\scriptsize -0.185 (-24.3\%)} & {\scriptsize -0.202 (-27.3\%)} \\
\cline{2-5}
& \textbf{RUS} & {\scriptsize -0.058 (-7.5\%)}  & {\scriptsize -0.138 (-18.1\%)} & {\scriptsize -0.141 (-19.0\%)} \\
\cline{2-5}
& \textbf{ES}  & {\scriptsize -0.094 (-12.2\%)} & {\scriptsize -0.155 (-20.3\%)} & {\scriptsize -0.157 (-21.2\%)} \\
\cline{2-5}
& \textbf{EN}  & {\scriptsize -0.049 (-6.35\%)} & {\scriptsize -0.142 (-18.63\%)} & {\scriptsize -0.15 (-20.2\%)} \\
\Xhline{1pt}

\multirow[c]{5}{*}{\textbf{RUS}} &
\textbf{-} & 0.613 & 0.638 & 0.641 \\
\cline{2-5}
& \textbf{FI}  & {\scriptsize -0.029 (-4.7\%)} & {\scriptsize -0.093 (-14.6\%)} & {\scriptsize -0.084 (-13.1\%)} \\
\cline{2-5}
& \textbf{AR}  & {\scriptsize -0.021 (-3.4\%)} & {\scriptsize -0.11 (-17.2\%)} & {\scriptsize -0.073 (-11.4\%)} \\
\cline{2-5}
& \textbf{ES} & {\scriptsize +0.012 (+1.95\%)} & {\scriptsize +0.0172 (+2.7\%)} & {\scriptsize +0.01 (+1.5\%)} \\
\cline{2-5}
& \textbf{EN} & {\scriptsize +0.022 (+3.48\%)} & {\scriptsize +0.019 (+2.97\%)} & {\scriptsize +0.013 (+2.0 \%)} \\
\Xhline{1.2pt}

\multirow[c]{5}{*}{\textbf{ES}} &
\textbf{-} & 0.437 & 0.468 & 0.457 \\
\cline{2-5}
& \textbf{FI} & {\scriptsize +0.073 (+16.7\%)} & {\scriptsize +0.036 (+7.7\%)} & {\scriptsize +0.029 (+6.3\%)} \\
\cline{2-5}
& \textbf{AR} & {\scriptsize +0.054 (+12.4\%)} & {\scriptsize +0.010 (+2.1\%)} & {\scriptsize +0.004 (+0.9\%)} \\
\cline{2-5}
& \textbf{RUS}  & {\scriptsize +0.06 (+13.72\%)} & {\scriptsize +0.041 (+0.87\%)} & {\scriptsize +0.063 (+13.7\%)} \\
\cline{2-5}
& \textbf{EN} & {\scriptsize +0.064 (+14.6\%)} & {\scriptsize +0.044 (+9.4\%)} & {\scriptsize +0.065 (+14.2\%)} \\

\Xhline{1pt}

\end{tabular}%
}
\caption{
Monolingual retrieval performance (nDCG@10) on MIRACL when fine-tuning \texttt{me5} on 150K target-language instances (baseline) and the relative change when adding 150K from another language. The threshold ($\tau$) critically affects how different language mixes influence performance.
}
\label{tab:monolingual_mixing_pairs_150k}
\end{table}

In multilingual dense retrieval, it is often desirable to leverage supervision from other languages, either to compensate for limited labeled data in the target language or to enrich the model’s representational space \citep[e.g.,][]{thakur-etal-2024-leveraging}.
We therefore extend the monolingual passage retrieval setting from the previous section to investigate how the relevance threshold behaves when fine-tuning on a mixture of languages. Specifically, we examine whether the optimal threshold depends solely on the target language or also on the particular combination of languages included in the fine-tuning data.

\subsection{Experimental Setup}

We adopt the same fine-tuning procedure described in Section~\ref{sec_monolingual}, but instead of training on a single language, we construct mixed fine-tuning datasets. For each target language, we select 150K instances from that language and shuffle them with an additional 150K instances from another language. A single relevance threshold is applied uniformly across both languages in the mixture.
All fine-tuned models are evaluated on monolingual passage retrieval using the MIRACL development set for the target language, with nDCG@10 as the primary metric.

\subsection{Results}

\autoref{tab:monolingual_mixing_pairs_150k} reports the results for each target language when fine-tuning with 150K instances from that language alone, along with the \emph{relative} performance changes observed when adding 150K instances from another language. Results are shown for thresholds $\tau\in\{1,2,3\}$.

We find that multilingual mixtures yield distinct behaviors depending on the target language, the enriching language, and the chosen threshold. Consistent with the monolingual setting, lower-resource target languages benefit from lower thresholds, even when the enrichment language is high-resource. In contrast, higher thresholds lead to increasing performance degradation.
For Arabic, adding fine-tuning data from any other language does not improve results and often worsens them, an effect that intensifies as the threshold increases.
Russian, in contrast, improves with higher thresholds and benefits comparably from high-resource enrichment languages across all thresholds. However, when the additional language is low-resource, performance drops, though less severely under a lower threshold.
Spanish exhibits a more nuanced pattern. When enriched with high-resource languages, gains are modest at threshold~2 and roughly comparable at thresholds~1 and~3. When enriched with Finnish or Arabic (both lower-resourced), however, improvements increase substantially as the threshold decreases, even though Spanish alone achieves stronger baseline performance when the threshold is higher.

These complex (and somewhat confusing) patterns emphasize that in multilingual mixture settings, a single global threshold is clearly suboptimal. The threshold appropriate for the target language may differ from that suitable for the enriching language, and mixing languages can shift the distribution of positive and negative examples in ways that interact complexly with language-specific properties.
Overall, the results underscore the importance of carefully selecting relevance thresholds, particularly in multilingual fine-tuning scenarios where both the target language and the composition of the training mixture jointly influence model behavior.

\section{Use-case 3: Crosslingual Retrieval}
\label{sec_crosslingual}

\begin{table}[t]
\centering
\small
\setlength{\tabcolsep}{3pt}
\renewcommand{\arraystretch}{1.15}

\resizebox{\columnwidth}{!}{
\begin{tabular}{|C{1.5cm}|C{1.2cm}|C{1.2cm}|C{1.2cm} !{\vrule width 1.2pt} C{1.2cm}|C{1.2cm}|C{1.2cm}|}
\hline

\cellcolor{white}\textbf{} &
\cellcolor{gray!20}\textbf{Corpus} &
\multicolumn{2}{c !{\vrule width 1.2pt}}{\cellcolor{gray!30}\textbf{Low-Med Resource}} &
\multicolumn{3}{c|}{\cellcolor{gray!30}\textbf{High Resource}} \\
\hline

\cellcolor{white}\textbf{Queries} &
\cellcolor{white}\textbf{} &
\cellcolor{gray!20}\textbf{AR} &
\cellcolor{gray!20}\textbf{JA} &
\cellcolor{gray!20}\textbf{RUS} &
\cellcolor{gray!20}\textbf{ES} &
\cellcolor{gray!20}\textbf{EN} \\
\hline

\multirow{2}{*}{\parbox{1.5cm}{\centering\textbf{Low-Med\\Resource}}} &
\cellcolor{white}\textbf{AR} &
\cellcolor{red!30}-2.4\% & \cellcolor{red!30}-15.58\% &
\cellcolor{red!30}-15.88\% & \cellcolor{red!30}-11.44\% & \cellcolor{red!30}-6.54\% \\
\cline{2-7}

& \cellcolor{white}\textbf{JA} &
\cellcolor{red!30}-17.71\% & \cellcolor{red!30}-4.2\% &
\cellcolor{red!30}-11.73\% & \cellcolor{red!30}-11.31\% & \cellcolor{red!30}-6.21\% \\

\Xhline{1.2pt}

\multirow{3}{*}{{\parbox{1.5cm}{\centering\textbf{High\\Resource}}}} &
\cellcolor{white}\textbf{RUS} &
\cellcolor{red!30}-14.93\% & \cellcolor{red!30}-10.03\% &
\cellcolor{green!30}+3.2\% & \cellcolor{green!30}+8.71\% & \cellcolor{green!30}+3.7\% \\
\cline{2-7}

& \cellcolor{white}\textbf{ES} &
\cellcolor{red!30}-17.71\% & \cellcolor{red!30}-8.36\% &
\cellcolor{green!30}+6.02\% & \cellcolor{green!30}+5.1\% & \cellcolor{green!30}+3\% \\
\cline{2-7}

& \cellcolor{white}\textbf{EN} &
\cellcolor{green!30}+0.90\% & \cellcolor{green!30}+5.15\% &
\cellcolor{green!30}+2.90\% & \cellcolor{green!30}+2.14\% & \cellcolor{green!30}+7.23\% \\
\hline

\end{tabular}
}
\caption{Relative change in cross-lingual retrieval performance when increasing the relevance threshold from 1 to 3 (negative values favor threshold~1, positive values favor threshold~3). Evaluated on the CLIRMatrix MULTI-8 benchmark. Rows correspond to query languages and columns to corpus languages; models are fine-tuned on the query language. Lower-resource languages generally prefer lower thresholds, while pairs of high-resource languages benefit from higher thresholds. See \autoref{fig:crosslingual} in the appendix for full results.}
\label{tab:crosslingual_perf}

\end{table}



In cross-lingual passage retrieval \citep[e.g.,][]{Litschko2018clir, Asai2021clpr}, the query is written in one language while relevant passages may appear in another. This scenario is common in multilingual search, where users prefer, or are only able, to formulate queries in their native language even when the corpus is in a different language. Cross-lingual retrieval aims to bridge this linguistic gap by enabling models to retrieve semantically relevant content across languages.

\subsection{Experimental Setup}

For this use case, we evaluate the same fine-tuned models from the monolingual experiments, specifically those trained with 300K instances, which was the more effective configuration. We test these models on the CLIRMatrix MULTI-8 benchmark \citep{sun2020clirmatrix}, which supports cross-lingual query-passage retrieval. Performance is measured using nDCG@10. The evaluation covers all language pairs for which both query and passage languages appear in the fine-tuning data, excluding Finnish, which is not available in this test set.

\subsection{Results}

First, our results show that fine-tuning generally improves cross-lingual retrieval performance when the relevance threshold is chosen appropriately. When the threshold is well matched to the language pair, the fine-tuned models consistently outperform the non-fine-tuned baseline; however, an unsuitable threshold often leads to underperformance relative to the baseline. These effects are visible in the curves in \autoref{fig:crosslingual} in the Appendix, which illustrate models' response to different thresholds across query-corpus combinations. 
Although more fine-tuning data might eventually mitigate failures caused by a poor threshold, choosing an appropriate threshold avoids most of these cases altogether.

\autoref{tab:crosslingual_perf} summarizes the plots, and quantifies these effects by presenting the relative performance differences when moving from threshold~1 to threshold~3. Negative values, in red, indicate that threshold~1 yields better performance, whereas positive values, in green, indicate that threshold~3 is more beneficial.

Overall, the patterns in \autoref{tab:crosslingual_perf} reveal systematic dependencies between threshold choice and language characteristics. When either the query language or the corpus language is lower-resource, lower thresholds tend to perform better, suggesting that preserving greater variability in relevance pairs is beneficial in such settings. English as a corpus language stands out as a notable exception, where higher thresholds can be advantageous, potentially owing to English’s dominance and extensive representation in pre-training corpora.

When both query \textit{and} corpus languages are high-resource (lower-right region of table), higher thresholds yield stronger results, reflecting already well-calibrated relevance distinctions in these languages.

Despite evaluating on CLIRMatrix rather than MIRACL, the broad trend aligns with the monolingual findings, with English again being the primary exception. These results reinforce that the degree of relevance variability between query and passage (operationalized through the threshold) must be chosen carefully with respect to both the language pair and the retrieval task.

\section{Concluding Discussion}
\label{sec_discussion}
Our analysis highlights that graded relevance annotations provide valuable nuance for training multilingual dense retrievers, but their effectiveness depends critically on how they are converted into positive and negative supervision. Because languages and retrieval tasks differ in data availability, semantic variation, and pre-training coverage, the optimal balance between strict and permissive relevance signals is not universal. Threshold calibration is therefore essential since it governs how much variability in relevance the model should internalize, enabling fine-tuning to adapt appropriately to different linguistic and task conditions.

We also observed notable disagreement between LLM annotators, especially in lower-resource languages. Annotation differences are another cause for controlled thresholding, but it also reinforces the need for mechanisms that can buffer against annotation noise. By controlling how graded labels are mapped to training signals, it can mitigate inconsistencies while still preserving the richer supervision signal that graded relevance offers.

Beyond these methodological considerations, our findings raise important questions about the role of pre-training and evaluation data. The extreme cross-lingual differences in the preferred variability of positive examples, and the unexpectedly high scores in several low-resource languages, indicate that multilingual retrievers may inherit uneven relevance structures shaped by their pre-training environments or by benchmark artifacts. Our analyses shed light on these dynamics, but systematic study of pre-training corpora, annotation behavior, and dataset construction is needed to fully understand and address these discrepancies.

Overall, our findings support our position that graded relevance with thoughtful thresholding is a powerful resource for multilingual retrieval. Calibration of the relevance boundary, rather than assuming a one-size-fits-all threshold, should be treated as a central methodological choice in the development of multilingual retrieval systems.




\section*{Limitations}

Our analysis covers several representative use cases in multilingual dense retrieval, but many real-world configurations remain unexplored. Retrieval tasks, model architectures, annotation pipelines, datasets, and languages can differ substantially, and we expect threshold behavior to vary accordingly. This diversity is precisely what we aim to highlight. Each setting requires its own assessment rather than relying on a single universal threshold.

Another consideration is that changing the relevance threshold alters the balance between positive and negative training instances. Although the groups remain large and our results show opposing effects across languages (suggesting that group size is not the dominant factor) this imbalance is inherent to real-world scenarios and should be taken into account when selecting a threshold.

Finally, our dataset construction involves retrieving candidate passages with a baseline model before annotation. Queries that are poorly matched to the corpus may produce more irrelevant candidate passages, introducing a potential bias in the resulting training data. This bias, which arises in any dataset prepared through similar pipelines, could influence fine-tuning behavior and warrants further study.


\bibliography{custom}

\appendix

\section{Implementation Details}
\label{app:implementation}
\paragraph{LLM annotation prompts.}
The first phase of preparing the synthetic dataset was generating queries for sampled passages from the MIRACL corpus. We prompted \texttt{GPT-4o} as follows:

\begin{tcolorbox}[colback=gray!10,colframe=gray!50]
\texttt{Your task is to write a <LANG> query that seeks information existing in a given context passage. \\
The query should be relevant to the passage but does not need to cover the entire scope of the passage. \\
The query should be in <LANG>, short, simple and in the style of web search query such as question-based or keyword-based.\\\\
The passage is:\\<PASSAGE>}
\end{tcolorbox}

An LLM (either \texttt{GPT-4o} or \texttt{Gemma-3-27B}) is later prompted to provide a relevance score for a query-passage pair. The prompt is:

\begin{tcolorbox}[colback=gray!10,colframe=gray!50]
\texttt{You are a search quality rater evaluating the relevance of passages. Given a query and a passage, you must provide a score on an integer scale of 0 to 3 within brackets [] with the following meanings:\\
{}[3] = Highly relevant: The passage is dedicated to the query and contains the exact answer.\\
{}[2] = Relevant: The passage provides a partial answer to the query, but it lacks the exact information needed, so it cannot be considered highly relevant.\\
{}[1] = Related: The passage seems related to the query but does not answer it.\\
{}[0] = Irrelevant: The passage has nothing to do with the query.\\\\
Query:\\<QUERY>\\\\
Passage:\\<PASSAGE>}
\end{tcolorbox}

\paragraph{LLM annotation cost.}
We used the Azure OpenAI API to use \texttt{GPT-4o}, and spent about \$3000 for the query generation phase and \$6000 for the relevance scoring annotations.
Annotating with \texttt{Gemma-3-27B} took several hours of GPU server time.

\paragraph{Downsampling the annotated instances.}
There were 500K query-passage pairs annotated, however to keep an approximately consistent distribution of relevance scores for the six languages, we downsampled to 300K as follows:
\begin{enumerate}
    \item The language with the smallest amount of non-zero scores (i.e., 1, 2 and 3) was identified.
    \item That number of instances was then sampled for the rest of the languages separately.
    \item For each language, instances with a score of 0 were sampled until 300K instances total were reached.
\end{enumerate}
As \autoref{fig:score_dist} shows, the distribution of scores are quite similar across languages.

\paragraph{Test set sizes.}
For the languages in our monolingual analyses in use-cases 1 and 2, the MIRACL dev set has between 648 and 2896 instances (queries) depending on the language (see: \url{https://huggingface.co/datasets/miracl/miracl}).
In the CLIRMatrix benchmark, used in the cross-lingual retrieval task (use-case 3), there are 1000 instances (queries) per language.

\paragraph{Fine-tuning the dense retrieval model.} We fine-tuned \texttt{multilingual-e5-large}, an encoder-based transformer, using the filtered $(q, p)$ pairs derived from our curated dataset, using the amount of instances that the experiment called for. Models were trained with a batch size of 32 using the InfoNCE loss with in-batch negatives, an initial learning rate of $1 \times 10^{-6}$, a 5\% warm-up ratio, and for two epochs. To ensure fair comparison across different threshold and data-scale configurations, we kept the training procedure fixed in terms of epochs, optimizer, and learning-rate schedule. While different thresholds lead to different loss behavior, our goal is to compare supervision settings under equal training budgets rather than to individually optimize each configuration.

The models were trained on a server with 24 CPUs, 220 GiB of RAM, and an 80GB NVIDIA A100 GPU.
A model took about 5 hours on average to finetune.

\paragraph{Use of AI.}
We used Copilot with GPT-4o as a backbone for small code improvements when coding the dataset creation pipeline. We also used Copilot to assist in preparing figures and tables for the paper, based on the results of our experiments.
Finally, we used ChatGPT 5.1 for assisting in paraphrasing some parts of the text while writing this paper to improve clarity and coherence.
Any use of AI was followed by manual verification and editing to maintain correctness and authenticity.

\section{Supplementing Results}
\label{app:supplementing_results}
\begin{figure*}[t!]
  \centering
  \includegraphics[width=\textwidth, , keepaspectratio]{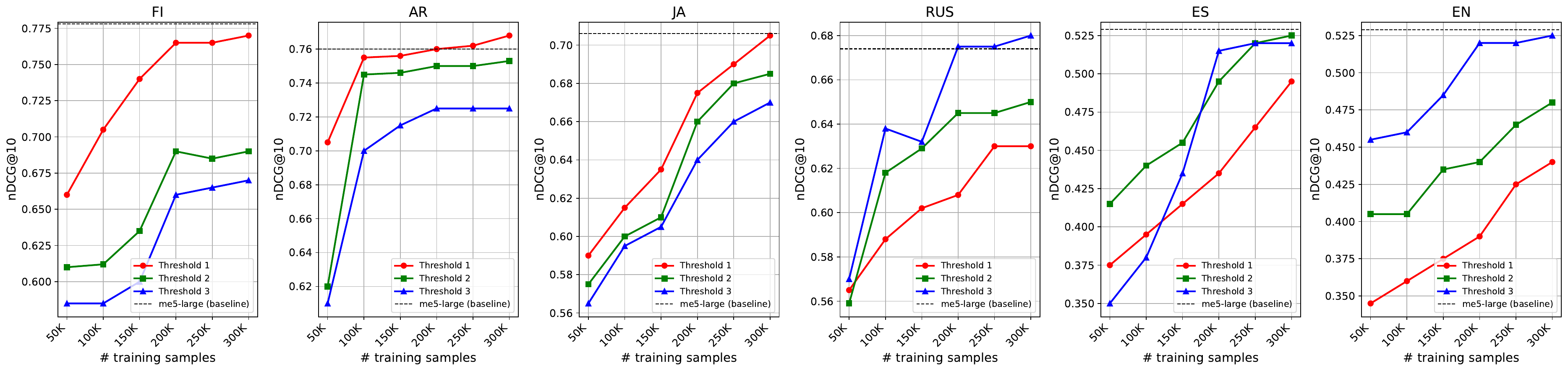}
  \caption{Monolingual retrieval performance (nDCG@10) of \texttt{me5} on MIRACL (horizontal dashed lines) and after fine-tuning, across six languages, when using the relevance scores annotated by \texttt{Gemma-3-27B}. Each subplot shows performance as a function of training set size under different relevance thresholds used to convert graded LLM scores into contrastive training pairs. Lower-resource languages tend to benefit from lower thresholds, whereas higher-resource languages perform better with higher thresholds. See \autoref{fig:mono_gpt} for the results when using \texttt{GPT-4o} as the annotator for relevance scores.}
  \label{fig:mono_gemma}
\end{figure*}


\begin{figure*}[t!]
  \centering
  \includegraphics[width=\textwidth]{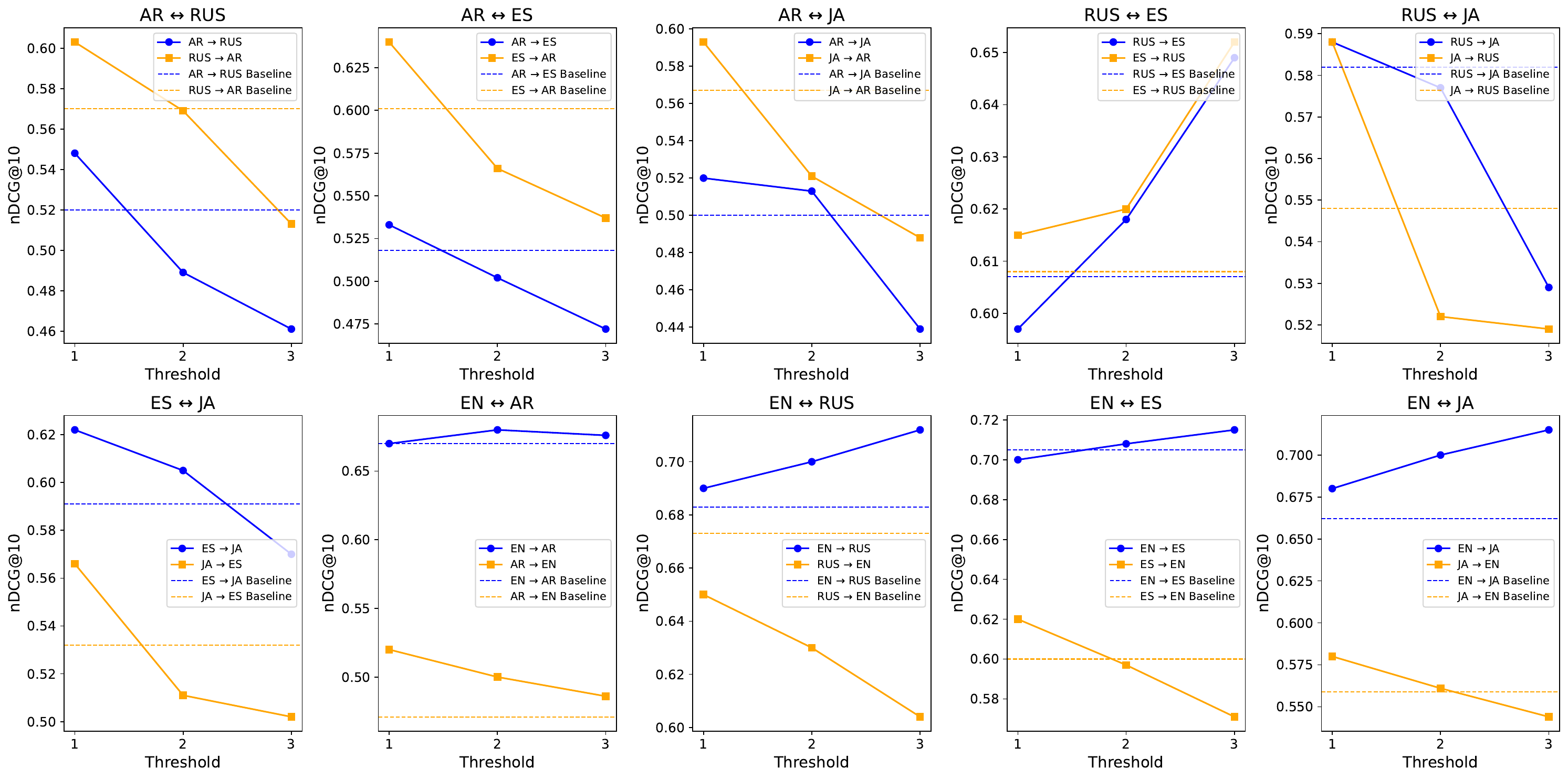}
  \caption{Performance of cross-lingual retrieval for various language pairs across thresholds (represented with <query-lang> → <corpus-lang>). Each subplot represents a language pair, showing nDCG@10 values for both retrieval directions, evaluated on the CLIRMatrix MULTI-8 benchmark. Dashed horizontal lines indicate the baseline performance achieved using \texttt{me5} for each direction. \autoref{tab:crosslingual_perf} summarizes the results.}
  \label{fig:crosslingual}
\end{figure*}


\paragraph{Use-case 1 (monolingual retrieval).}
\autoref{fig:mono_gemma} shows the results on the monolingual task (use-case 1, \S\ref{sec_monolingual}) when \texttt{Gemma-3-27B} was used as the annotator for relevance scores. This figure is to be compared to \autoref{fig:mono_gpt} in which \texttt{GPT-4o} was used as the annotator instead. As described in Section \ref{sec_monolingual}, the performance when using \texttt{Gemma} as the annotator is not as good, but trends are similar with a few differences. This is another example of the importance of setting the threshold sensibly with respect to language, data and task.

\paragraph{Use-case 3 (cross-lingual retrieval).}
\autoref{fig:crosslingual} presents results for the cross-lingual experiments. Each plot in the figure reports the performance for both retrieval directions (X→Y and Y→X; represented as <query-lang> → <corpus-lang>), with baseline results from the non–fine-tuned \texttt{multilingual-e5-large} model shown for reference. The figure is also summarized in \autoref{tab:crosslingual_perf}. See discussion in Section \ref{sec_crosslingual}.


\end{document}